\documentclass{ws-jbcb}
\newcommand{\titel}{Uniqueness, intractability and exact algorithms:\\reflections on level-k phylogenetic networks}
\usepackage{makeidx}
\usepackage{amsfonts}
\usepackage[letterpaper,hmargin=1.15in,vmargin=1in]{geometry}
\usepackage[british]{babel}
\usepackage{subfigure}
\usepackage{graphicx}
\usepackage[pdftex, colorlinks=true, linkcolor=blue, citecolor=blue, urlcolor=blue, hypertexnames=true, letterpaper=true, pdfstartview=FitH]{hyperref}
\hypersetup{ %
pdfauthor = {Leo van Iersel, Steven Kelk and Matthias Mnich}, %
pdftitle = {\titel}, %
pdfsubject  = {Phylogenetic Network Construction}, %
pdfkeywords = {exact algorithm, phylogenetic networks, triplets, NP-hard}, %
}

\begin{document}


%
\catchline{}{}{}{}{}
%

\title{\titel\thanks{Part of this research has been funded by the Dutch BSIK/BRICKS project.}}

\author{Leo van Iersel}
\address{Department of Mathematics and Computer Science, Technische Universiteit Eindhoven,\\P.O. Box 513, 5600 MB Eindhoven, The Netherlands\\\href{mailto:l.j.j.v.iersel@tue.nl}{{l.j.j.v.iersel@tue.nl}}}

\author{Steven Kelk}
\address{Centrum voor Wiskunde en Informatica (CWI),\\P.O. Box 94079, 1090 GB Amsterdam, The Netherlands\\ \href{mailto:s.m.kelk@cwi.nl}{{s.m.kelk@cwi.nl}}}

\author{Matthias Mnich}
\address{Department of Mathematics and Computer Science, Technische Universiteit Eindhoven,\\P.O. Box 513, 5600 MB Eindhoven, The Netherlands\\\href{mailto:m.mnich@tue.nl}{{m.mnich@tue.nl}}}

\maketitle

\begin{history}
\end{history}

\begin{abstract}
Phylogenetic networks provide a way to describe and visualize evolutionary histories that have undergone so-called
reticulate evolutionary events such as recombination, hybridization or horizontal gene transfer. The level $k$ of a
network determines how non-treelike the evolution can be, with level-0 networks being trees. We study the problem of
constructing level-$k$ phylogenetic networks from triplets, i.e. phylogenetic trees for three leaves (taxa). We give,
for each $k$, a level-$k$ network that is uniquely defined by its triplets. We demonstrate the applicability of this
result by using it to prove that (1) for all $k \geq 1$ it is NP-hard to construct a level-$k$ network consistent with
all input triplets, and (2) for all $k \geq 0$ it is NP-hard to construct a level-$k$ network consistent with a
maximum number of input triplets, even when the input is dense. As a response to this intractability we give an exact
algorithm for constructing level-1 networks consistent with a maximum number of input triplets.
\end{abstract}


\section{Introduction}
A central problem in biology is to accurately reconstruct plausible evolutionary histories. This area of research is
called phylogenetics and provides fascinating challenges for both biologists and mathematicians. Throughout most of
the history of phylogenetics researchers have concentrated on constructing phylogenetic \emph{trees}. In recent years
however, more and more attention is devoted to phylogenetic \emph{networks}. From a biological point of view, networks
are able to explain and visualize more complex evolutionary scenarios, since they take into account biological
phenomena that cannot be displayed in a tree. These phenomena are so-called reticulate evolutionary events such as
hybridization, recombination and horizontal gene transfer. From a mathematical point of view however, phylogenetic
networks pose formidable challenges. Irrespective of the exact model being used, many problems that are
computationally tractable for trees (i.e. solvable in polynomial time) become intractable (NP-hard) for networks.
Huson and Bryant wrote a detailed discussion of phylogenetic networks and their application \cite{HusonBryant2006}.
Here we study the \emph{level} of networks, which restricts how interwoven the reticulations can be. In trees (i.e.
level-0 networks) no reticulation events occur; in level-1 networks all reticulation cycles must be disjoint. The
higher the level of the network, the more freedom in reticulation is allowed. Formally, a level-$k$ network is a
phylogenetic network in which each biconnected component contains at most $k$ reticulation events. Level-1 networks
have also been called \emph{galled trees} \cite{GusfieldEtAl2004}, \emph{gt-networks} \cite{NakhlehEtAl2005} and
\emph{galled networks} \cite{JanssonEtAl2006}. General level-$k$ networks were first introduced by Choy, Jansson,
Sadakane and Sung \cite{ChoyEtAl2005}. The focus on level (as opposed to, for example, minimizing the total number of
reticulation vertices) is motivated by several factors. Firstly, level induces a hierarchy on the space of networks
with lower level networks being more 'tree-like' than higher-level networks. Identifying the position of candidate
solutions (i.e. networks) within this hierarchy, or finding the minimum level at which candidate solutions exist,
communicates important structural information about the solution space. (Level minimization, which derives its
legitimacy from the parsimony principle, can also be used in an implicit context e.g. to measure the accuracy of input
data. For example, if we expect the solution to be a tree, but only obtain higher level networks, this suggests that
data errors lie in the regions corresponding to the biconnected components.) Secondly, from an
algorithmic/mathematical perspective focussing on lower-level networks can yield corresponding improvements in
tractability/running time and to clearer mathematical analysis. Finally, restricting level is for many optimization
criteria necessary to avoid trivial solutions, e.g. several of the problems we discuss in this article can be
trivially optimized if we choose a solution with high enough level, but (as we shall see) this communicates no useful
information.

A great variety of approaches have been proposed for phylogenetic reconstruction. They include methods like Maximum
Parsimony, Maximum Likelihood, quartet-based methods, Bayesian methods (using Monte Carlo Markov Chain),
distance-based methods (using e.g. Neighbor Joining or UPGMA) and many others. They all have their advantages and
drawbacks \cite{BryantSteel2001,HolderLewis2003,JiangEtAl2001,SempleSteel2003}.

In this article we consider a triplet-based approach to construct directed phylogenetic networks. As input we take a
collection of triplets, which are rooted phylogenetic trees on size-3 subsets of the taxa. These triplets can, for
example, be constructed by methods such as Maximum Parsimony or Maximum Likelihood, that work accurately and fast for
small numbers of taxa. Another possibility is to infer the triplets from a set of phylogenetic trees, possibly
originating from different sources. However the triplets are obtained, the next step is to combine them into a single,
large phylogenetic network for all taxa. Designing algorithms for the latter task forms the subject of this article.
Triplet methods have become popular since they allow us to solve certain problems in polynomial time, as will be
elaborated on shortly. Next to that, an advantage of these methods is that they provide the possibility to combine
different sorts of biological data.

Triplet-based methods have been extensively studied in the literature. Aho et al. \cite{AhoEtAl1981} gave a
polynomial-time algorithm that constructs a tree from triplets if there exists a tree that is consistent with all
input triplets. This positive result provided the stimulus for studying the applicability of triplet-based methods to
networks. Unfortunately, it has been shown that for level-1 \cite{JanssonEtAl2006} and level-2
\cite{VanIerselEtAl2008,VanIerselEtAl2007} networks the corresponding problem becomes NP-hard. However, the same
articles give polynomial-time algorithms for the problem where the input is \emph{dense}, i.e. there is at least one
triplet in the input for every size-3 subset of the taxa. A related problem that accommodates errors in the triplets
is finding a tree consistent with as many input triplets as possible. This problem is NP-hard
\cite{Bryant1997,Jansson2001,Wu2004}, and approximation algorithms have been explored both for the construction of
trees \cite{Gasieniec1999} and level-1 networks \cite{ByrkaEtAl2008,JanssonEtAl2006}. For the construction of trees,
efficient heuristics have been designed by Semple and Steel \cite{SempleSteel2000}, Page \cite{Page2002}, Wu
\cite{Wu2004} and Snir and Rao \cite{SnirRao2006}. The last algorithm (MAX CUT triplets) outperforms the
character-based method Matrix Representation with Parsimony (MRP), which is popular in practice
\cite{Baum1992,Ragan1992,Sanderson1998}.

In this paper we study the structure and construction of level-$k$
networks. First, we analyze the minimum level $k$ ensuring that for
each input triplet set on $n$ leaves there exists a level-$k$
network consistent with all triplets (in Sect.~\ref{sec:suf}). Then
we use this analysis to give, for each $k$, a level-$k$ network that
is uniquely defined by the set of triplets it is consistent with,
i.e. no other level-$k$ network is consistent with that set of
triplets (in Sect.~\ref{sec:unique}). These networks we use to give
two NP-hardness results in Sect.~\ref{sec:nphard}. We prove that
constructing level-$k$ networks consistent with all input triplets
is NP-hard for every $k\geq 1$. This complements the known results
for $k \in \{1,2\}$ (see above), of which our result for $k > 2$ is
a non-trivial generalization. In addition, we show that constructing
a level-$k$ network consistent with a maximum number of input
triplets is NP-hard for all $k\geq 0$, even if the input triplet set
is dense. This means that it is even NP-hard to construct a
phylogenetic tree consistent with a maximum number of triplets from
a \emph{dense} triplet set. We respond to the aforementioned
intractability results with an exact algorithm for constructing
level-1 phylogenetic networks in Sect.~\ref{sec:exact}. This
algorithm runs in time $O(m4^n)$ (for $n$ leaves and $m$ triplets)
and can also be used for the weighted version of the problem.
Authors working on the unrooted analogue of triplets,
\emph{quartets}, have noted that their methods are particularly
powerful when the input quartets are chosen carefully (and are, for
example, not forced to contain information for each quadruple of
leaves) \cite{SnirEtAl2008}. The level-1 algorithm we present can
tolerate such inputs (i.e. non-dense input sets) and for this reason
we are optimistic about the biological relevance of the solutions it
produces. We conclude with a discussion of open problems.

\section{Preliminaries}
A \emph{phylogenetic network} (\emph{network} for short) is defined as a directed acyclic graph in which a single
vertex has indegree 0 and outdegree 2 (the \emph{root}) and all other vertices have either indegree 1 and outdegree 2
(\emph{split vertices}), indegree 2 and outdegree 1 (\emph{reticulation vertices}) or indegree 1 and outdegree 0 (\emph{leaves}), where the leaves are distinctly labeled.
Let $L(N)$ denote the set of leaves of a network $N$.

A directed acyclic graph is \emph{connected} (also ``weakly
connected'') if there is an undirected path between any two
vertices, and \emph{biconnected} if it contains no vertex whose
removal disconnects the graph. A \emph{biconnected component} of a
network is a maximal biconnected subgraph and is called
\emph{trivial} if it is equal to two vertices connected by an arc.
Otherwise, it is \emph{non-trivial}. An arc $a = (u,v)$ of a network
$N$ is a \emph{cut-arc} if its removal disconnects $N$; it is
\emph{trivial} if $v$ is a leaf and otherwise \emph{non-trivial}. A
vertex $w$ is \emph{below} an arc $a = (u,v)$ (and \emph{below}
vertex $v$) if there is a directed path from $v$ to $w$.

\begin{definition}
A network is said to be a \emph{level-$k$ network} if each biconnected component contains at most $k$ reticulation vertices.
\end{definition}

To avoid ``redundant'' networks, we require every non-trivial biconnected component of a network to have at least three outgoing arcs.
A level-$k$ network is a \emph{strict} level-$k$ network if it is not a level-$(k-1)$ network.
The class of level-0 networks are \emph{phylogenetic trees} (\emph{trees} for short); they have no reticulation vertices.

\begin{figure}
  \centering
  \subfigure[]
  {
    \centering
    \hspace{1.7cm}
    \includegraphics[scale=0.8]{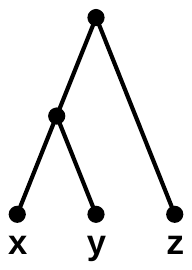}
    \hspace{1.7cm}
    \vspace{-0.5cm}
    \label{fig:singletriplet}
  }
  \subfigure[]
  {
    \centering
    \includegraphics[scale=0.8]{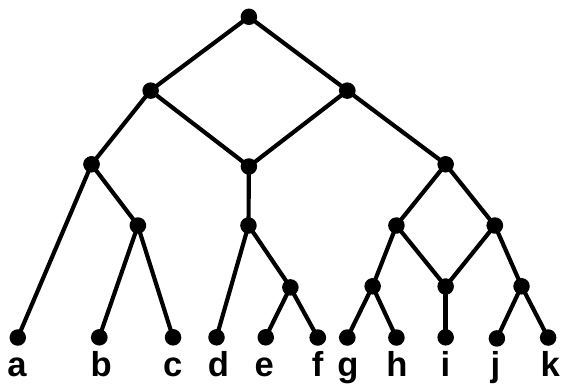}
    \vspace{-.1cm}
    \label{fig:bignetwork}
  }\
\caption{(a) One of three possible triplets on leaves $x,y,z$ and (b) an example of a level-1 network. As with all
figures in this article, all arcs are directed downwards.}
\end{figure}

A \emph{triplet} $xy|z$ is a tree on leaves $x,y,z$ such that the lowest common ancestor of $x$ and $y$ is a proper descendant of the lowest common ancestor of $x$ and $z$, see Fig.~\ref{fig:singletriplet}.
The leaves of triplet $t$ form the set $L(t)$.
A set $T$ of triplets has leaf set $L(T)=\bigcup_{t\in T}L(t)$, with size $n = L(T)$.
For $L' \subseteq L(T)$ denote by $T|_{L'}$ the set of triplets $t\in T$ with $L(t)\subseteq L'$.
A set $T$ of triplets is \emph{dense} if it contains at least one triplet for each size-3 subset of $L(T)$.

\begin{definition}
\label{def:con}
A triplet $xy|z$ is \emph{consistent} with a network $N$ (interchangeably: $N$ is consistent with $xy|z$) if $N$ contains a subdivision of $xy|z$,
i.e. if $N$ contains vertices $u \neq v$ and pairwise internally vertex-disjoint paths $u \rightarrow x$, $u \rightarrow y$, $v \rightarrow u$ and $v \rightarrow z$.
\end{definition}

By extension, a set $T$ of triplets is \emph{consistent} with $N$ (interchangeably: $N$ is consistent with $T$) if every triplet in $T$ is consistent with $N$ and $L(T) = L(N)$.
For example, Fig.~\ref{fig:bignetwork} is a level-1 network with two non-trivial biconnected components, each of them containing one reticulation vertex.
This network is consistent with (amongst others) the triplets $bc|a$, $bd|h$, $hd|b$ and $gi|k$, but is not consistent with $dg|k$, $ab|c$, $gk|i$ or $cd|f$.

\begin{figure}
  \centering
  \vspace{0.2cm}
  \includegraphics[width=0.6\textwidth]{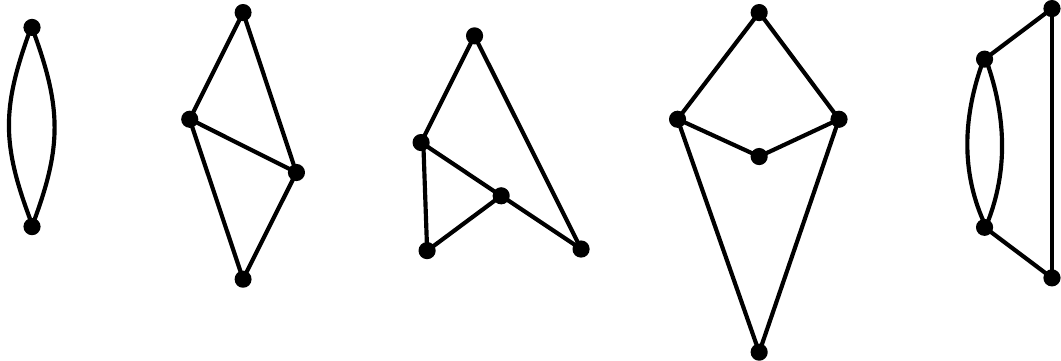}
  \caption{The unique simple level-1 generator and the four simple level-2 generators.}
  \label{fig:simplegen}
\end{figure}

We introduce the class of \emph{simple} level-$k$ networks.
Intuitively, these are the basic building blocks of level-$k$
networks in the sense that each non-trivial biconnected component of
a level-$k$ network is in essence a simple level-$l$ network, for
some $l\leq k$. These simple networks will be built by adding leaves
to ``generators'', which are formally defined as follows.
\begin{definition}
A \emph{simple level-$k$ generator}, for $k\geq 1$, is a directed
acyclic biconnected multigraph, which has a single root (a vertex
with indegree 0 and outdegree 2), precisely $k$ reticulation
vertices (with indegree 2 and outdegree at most 1) and apart from
that only split vertices (with indegree 1 and outdegree 2).
\end{definition}
A case analysis shows that there is only one simple level-1 generator and that there are four simple level-2 generators \cite{VanIerselEtAl2007}, depicted in Fig.~\ref{fig:simplegen}.
Computer calculations revealed the 65 simple level-3 generators \cite{Kelk2008}.

\begin{definition}
A simple level-$k$ network, for $k\geq 1$, is a network obtained by
applying the following transformation to some simple level-$k$
generator $G$:
\begin{enumerate}
\item first, for each pair $u,v$ of vertices in $G$ connected by a single arc $(u,v)$, replace $(u,v)$ by a path
with $\ell \geq 0$ internal vertices and for each such internal vertex $w$ add a new leaf $x$ and an arc $(w,x)$;
\item second, for each pair $u,v$ of vertices in $G$ connected by multiple arcs replace one such arc by a path with
at least one internal vertex and for each such internal vertex $w$
add a new leaf $x$ and an arc $(v,x)$; and treat the other arc
between $u,v$ as in step (1); \item third, for each vertex $v$ of
$G$ with indegree 2 and outdegree 0 add a new leaf $y$ and an arc
$(v,y)$.
\end{enumerate}
\end{definition}

We remark that at least three leaves have to be added to $G$, to
avoid redundancy of the constructed network. A network is
\emph{simple} if it is a simple level-$k$ network for some $k$.
There is an elegant characterisation of simple level-$k$ networks:
\begin{lemma}[Van Iersel et al. \cite{VanIerselEtAl2007}]
\label{lem:withoutcut}
For $k\geq 1$, a network $N$ is a simple level-$k$ network if and only if $N$ is a strict level-$k$ network and every cut-arc is trivial.
\end{lemma}

In our proofs we will frequently remove leaves from a network. This
might result in a graph that is not a valid network. Therefore, we
define \emph{tidying up} a directed acyclic graph as repeatedly
applying the following four steps: (1) delete unlabeled vertices
with outdegree 0; (2) suppress vertices with indegree and outdegree
1; (3) replace multiple arcs by single arcs and (4) replace
nontrivial biconnected components with at most two outgoing arcs by
a single vertex. Observe that if $N'$ is the result of removing
leaves $L'$ from network $N$ and tidying up the resulting graph,
then $N'$ is a valid network. In addition, observe that in this case
$N'$ is consistent with exactly the same triplets as $N$ is, except
for triplets containing leaves from $L'$.

\section{Sufficiency and Necessity of Network Level}
\label{sec:suf} In this section we prove that any triplet set on $n$ leaves is consistent with a level-$(n - 1)$
network. Then we show that this bound is tight by giving a triplet set on $n$ leaves that is not consistent with any
network of level smaller than $n-1$.

Let $T_F(n)$ be the set of all $3\binom{n}{3}$ triplets possible on $n$ leaves.
Call $T_F(n)$ the \emph{full triplet set} on $n$ leaves.

\begin{proposition}
For any triplet set $T$ on $n$ leaves there exists a level-$(n-1)$ network consistent with $T$.
\end{proposition}
\begin{proof}
Let $n \geq 3$ and let $N_F(n)$ be the network in Fig.~\ref{fig:fulltripletset}.
\begin{figure}
  \centering
  \includegraphics[scale=0.8]{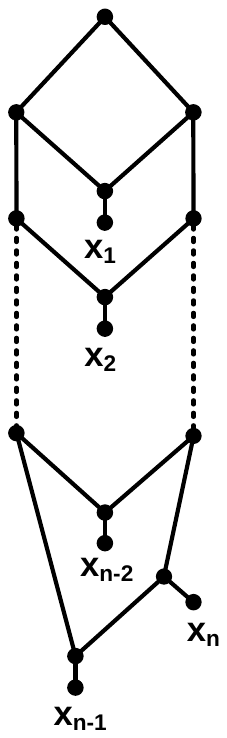}
  \caption{The level-$(n-1)$ network $N_F(n)$ is consistent with every triplet set on $n$ leaves, and has the minimum number of arcs and vertices among all such networks.}
  \label{fig:fulltripletset}
\end{figure}
First, look at triplets $x_hx_i|x_j \in T_F(n)$ with $h,i \neq n$.
There exists a unique split vertex $v$ below the left child of the
root, from which there are two paths to $x_h$ and $x_i$ that have
only $v$ in common. On the other hand, there is a path from the root
to $x_j$, via the right child of the root. So the network is
consistent with $x_hx_i|x_j$.

Second, look at triplets $x_ix_n|x_j \in T_F(n)$. There exists a
unique split vertex $v$ below the right child of the root, from
which there are two paths to $x_i$ and $x_n$ that have only $v$ in
common. As there is also a path from the root to $x_j$ via the left
child of the root, the network is consistent with $x_ix_n|x_j$.
Given that $T \subseteq T_F(n)$, the result follows.
\end{proof}

\begin{lemma}
\label{lem:issimple}
Any network consistent with the full triplet set must be simple.
\end{lemma}
\begin{proof}
Let $n\geq 3$ and let $N$ be consistent with $T_F(n)$.
If $N$ is not simple then, by Lemma~\ref{lem:withoutcut}, it contains a non-trivial cut-arc $a = (u,v)$.
If there is only one leaf below $a$, then $N$ is not a valid network because it contains a biconnected component with only one outgoing arc, which we do not allow.
If all leaves are below $a$, then again $N$ is not a valid network because it contains a biconnected component with only one outgoing arc.
Hence there are leaves $x$ and $y$ below $a$ and a leaf $z$ not below $a$.
This implies that the triplet $xz|y$ is not consistent with $N$, a contradiction.
\end{proof}

\noindent Let a \emph{reticulation leaf} be defined as a leaf whose parent is a reticulation vertex. Simple level-$k$
networks have, by definition, at least one and at most $k$ reticulation leaves ($k\geq 1$).

\begin{proposition}
\label{thm:fulltripletset}
The full triplet set on $n\geq 3$ leaves is not consistent with any level-$k$ network with $k < n - 1$.
\end{proposition}
\begin{proof}
The proof is by induction on $n$. The theorem holds for $n = 3$: any
network consistent with $T_F(3)$ must be simple by
Lemma~\ref{lem:issimple}, and any simple level-1 network on three
leaves is consistent with only two triplets. Since there are three
possible triplets on a set of three leaves, there is no level-1
network consistent with $T_F(3)$. Let $n > 3$; the induction
hypothesis is that for all $n' < n$ the full triplet set $T_F(n')$
is not consistent with any level-$k'$ network for $k' < n' - 1$.
Suppose for contradiction that the theorem does not hold for $n$,
thence there exists a level-$k$ network $N$ consistent with $T_F(n)$
and $k < n-1$. By Lemma~\ref{lem:issimple}, $N$ must be a simple
level-$k$ network and thus contains a reticulation leaf $x$. Delete
$x$ and tidy up the resulting graph. This decreases the level of the
network since the parent of $x$ is a reticulation vertex and gets
removed when tidying up the graph. This thus yields a level-$(n-3)$
network consistent with $T_F(n-1)$, contradicting the induction
hypothesis. We thus conclude that there exists no level-$k$ network
consistent with $T_F(n)$ for $k < n-1$.
\end{proof}

The network $N_F(n)$ of Fig. \ref{fig:fulltripletset} is much smaller than the network proposed by Jansson and Sung \cite{JanssonSung2006}, that is consistent with $T_F(n)$ and was obtained from a complicated sorting network.
\begin{lemma}
For $n\geq 3$, the network $N_F(n)$ has the minimum number of arcs and vertices over all networks consistent with the
full triplet set $T_F(n)$.
\end{lemma}
\begin{proof}
We first show that any simple level-$k$ network $N=(V,A)$ on $n$
leaves has $2n+2k-1$ vertices and $2n+3k-2$ arcs. Let $s$ be the
number of split vertices. The sum of the indegrees of all vertices
is $s+2k+n$, while the sum of their outdegrees is $2+2s+k$. It is
well known that in any directed graph the sum of all outdegrees
equals the sum of all indegrees. It follows that $s=n+k-2$. Using
this formula we obtain that the total number of vertices equals:
\begin{equation*}
|V| = s + k + n + 1 = (n + k - 2) + k + n + 1 = 2n + 2k - 1\enspace .
\end{equation*}
Split vertices and reticulation vertices have total degree 3, leaves
have total degree 1, and the root of $N$ has total degree 2. Thus
the total number of arcs in $N$ is:
\begin{equation*}
|A| = \frac{3s + 3k + n + 2}{2} = \frac{3(n+k-2) + 3k + n + 2}{2} = 2n + 3k - 2\enspace .
\end{equation*}

Let $N_n$ be a network consistent with $T_F(n)$. By Lemma~\ref{lem:issimple} and Proposition~\ref{thm:fulltripletset},
$N_n$ is simple and has level at least $n-1$. Then the calculation above yields that $N_n$ has at least $2n+2(n-1)-1 =
4n-3$ vertices and $2n+3(n-1)-2 = 5(n-1)$ arcs. The proof is complete by noting that $N_F(n)$ has exactly $4n-3$
vertices and $5(n-1)$ arcs.
\end{proof}

\section{A Unique Level-$k$ Network}
\label{sec:unique} In the construction and analysis of triplet methods, it is often important to know that a certain
network is uniquely defined by a set of triplets. Characterizing such networks is an important open problem. In this
section we present a partial solution to this question by giving, for each $k$, a level-$k$ network $N^k$ that is
unique in the sense that it is the only level-$k$ network that is consistent with all triplets that are consistent
with $N^k$. In the next section we will demonstrate how useful this unique network is, by using it to show the
intractability of constructing level-$k$ networks from triplets.

\begin{figure}
  \centering
  \includegraphics[width=\textwidth]{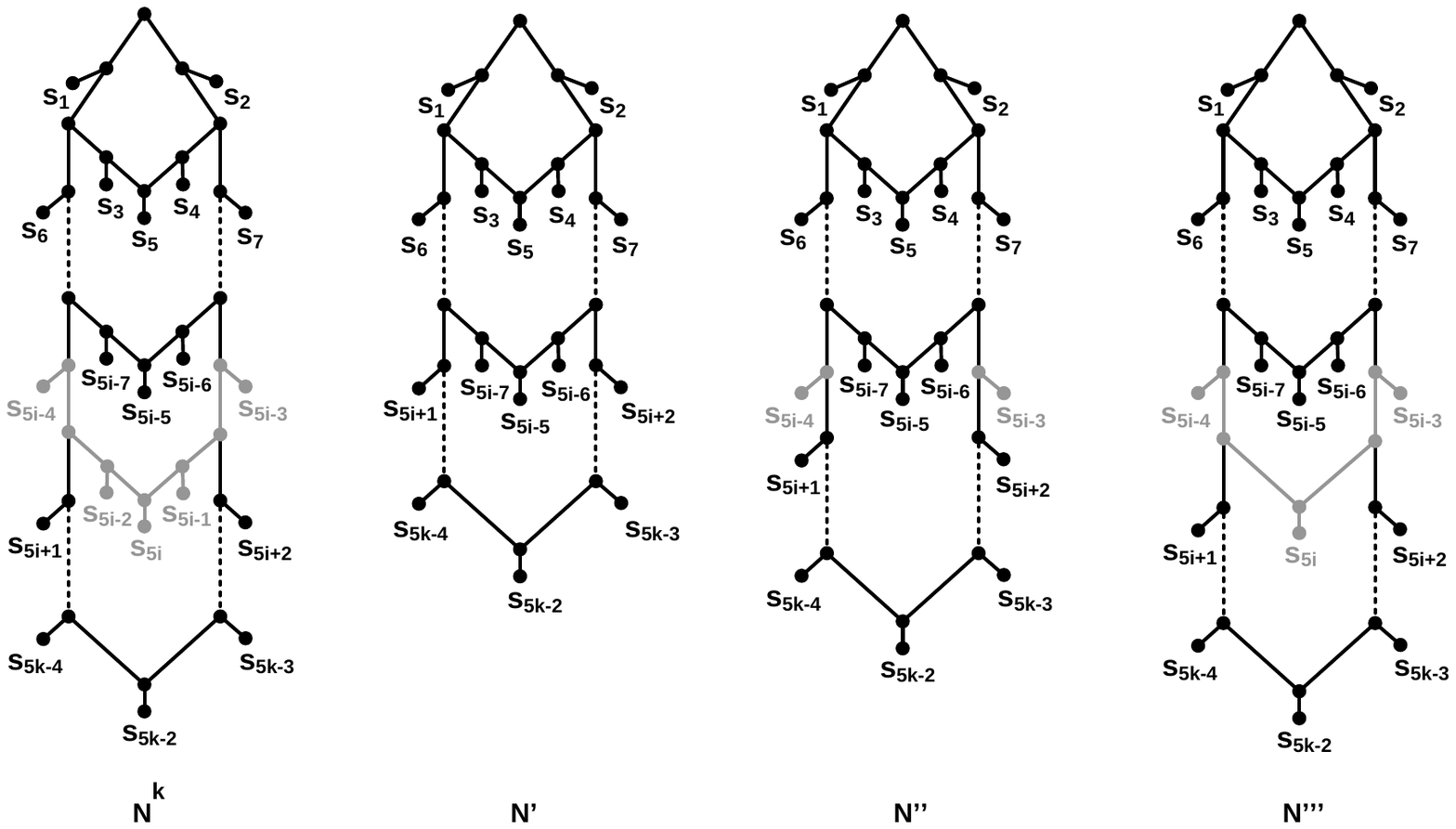}
  \caption{Reconstruction of $N^k$ from $N'$.
           After adding $s_{5i-4}$ and $s_{5i-3}$ to $N'$ we obtain network $N''$.
           After also adding $s_{5i}$ we obtain $N'''$.
           Finally, after adding $s_{5i-2}$ and $s_{5i-1}$ we obtain the original network $N^k$ to the left again.}
  \label{fig:levelkladder}
\end{figure}

Let $N^k$ be the network to the left in Fig.~\ref{fig:levelkladder} and let $T^k$ be the set of triplets that are
consistent with $N^k$. By \emph{hanging leaves} $x_1,\hdots,x_p$ on an arc $(u,v)$ (for some $p\geq 1$) we mean
replacing $(u,v)$ by a path $u,w_1,\hdots,w_p,v$ and adding arcs $(w_i,x_i)$ for all $i=1,\hdots,p$.

\begin{theorem}
\label{thm:unique}
For each $k\geq 2$, the network $N^k$ is the unique level-$k$ network consistent with $T^k$.
\end{theorem}
\begin{proof}
Let $R$ be the set of reticulation leaves of $N^k$, that is $R=\{s_5,s_{10},\hdots,s_{5k-5},s_{5k-2}\}$.
We start by proving the following claims.
\begin{claim}
\label{claim:claim1}
Any level-$k$ network consistent with $T^k$ is a simple level-$k$ network.
\end{claim}
\begin{proof}[Proof of Claim~1]
First observe that all triplets over the leaves $R\cup\{s_{5k-4}\}$ are in $T^k$.
Let $N$ be a level-$k$ network consistent with $T^k$.
From Proposition~\ref{thm:fulltripletset} it follows that $N$ is a strict level-$k$ network.
Now suppose for contradiction that $N$ is not simple.
Then by Lemma~\ref{lem:withoutcut}, $N$ contains a non-trivial cut-arc $a$.
Let $B\subseteq L(N)$ be the set of leaves below $a$ and let $A = L(N) \setminus B$.
Because $a$ is non-trivial, $B$ contains at least two leaves.
For every two leaves $x,y$ in $B$ and every leaf $z$ in $A$, there is only one triplet in $T^k$ on leaves $x,y,z$ that is consistent with $N$.
However, for $s_{5k-2}$ there are no two leaves $x',y'$ such that there is only one triplet in $T^k$ with leaves $s_{5k-2},x',y'$.
It follows that $s_{5k-2}$ belongs to neither $A$ nor $B$, a contradiction.
\end{proof}

\begin{claim}
\label{claim:claim2}
In any level-$k$ network consistent with $T^k$, at least one of the leaves in $R$ is a reticulation leaf.
\end{claim}
\begin{proof}[Proof of Claim~2]
Let $N$ be a network consistent with $T^k$. Recall that $T^k$
contains all possible triplets over leaves $R\cup\{s_{5k-4}\}$.
Proposition~\ref{thm:fulltripletset} says that any network
consistent with all triplets over $R\cup\{s_{5k-4}\}$ cannot have
level smaller than $k$, so $N$ is a strict level-$k$ network. By
Claim~1, $N$ is a simple level-$k$ network and hence contains a
reticulation leaf $x$. First suppose $x$ does not belong to
$R\cup\{s_{5k-4}\}$. Then removing $x$ and tidying up the resulting
graph yields a level-$(k-1)$ network consistent with all triplets
over $R\cup\{s_{5k-4}\}$. A contradiction, thus $N$ contains no
leaves outside $R\cup\{s_{5k-4}\}$ as reticulation leaf.
Symmetrically, no leaf outside $R\cup\{s_{5k-3}\}$ is a reticulation
leaf of $N$. It follows that only leaves from $R$ can be
reticulation leaves of $N$, so $x$ belongs to $R$.
\end{proof}

\noindent We are now ready to prove the theorem. The proof is by induction on $k$; the base case $k = 2$ has been
proven by Van Iersel et al.~\cite[Lemma~18]{VanIerselEtAl2007}. Let $k > 2$ and assume the theorem holds for all $k' =
2,\hdots,k-1$. In the induction step, we will show that any level-$k$ network consistent with $T^k$ and with
reticulation leaf $s_{5i}$ (for any $i\in\{1,\ldots, k-1\}$), equals the network $N^k$. The case that $s_{5k-2}$ is a
reticulation leaf is symmetric to the case that $s_{5k-5}$ is a reticulation leaf. Since by Claim~\ref{claim:claim2}
at least one leaf from $R$ must be a reticulation leaf, the theorem will follow.

Let $N$ be a simple level-$k$ network consistent with $T^k$ and with
reticulation leaf $s_{5i}$ (with $i\in\{1,\ldots,k-1\}$). Let $T'$
be the triplet set obtained from $T^k$ by removing all triplets
containing some leaf from $\{s_{5i-4},\hdots,s_{5i}\}$, i.e. $T' =
T^k|_{(L\setminus\{s_{5i-4},\hdots,s_{5i}\})}$. Then $T'$ is
consistent with network $N'$, the second network from the left in
Fig.~\ref{fig:levelkladder}. Because $T'$ equals the set of all
triplets that are consistent with $N'$ (which is a relabeling of
$N^{k-1}$), by the induction hypothesis $N'$ is the unique
level-$(k-1)$ network consistent with $T'$.

Consider the network obtained from $N$ by removing the leaves
$s_{5i-1},s_{5i-2},s_{5i},s_{5i-3}$ and $s_{5i-4}$ (in this order)
from $N$ and tidying up the resulting graph. This decreases the
level of the network, since the parent of $s_{5i}$ was a
reticulation vertex and gets removed when tidying up the graph.
Hence this gives a level-$(k-1)$ network consistent with $T'$, which
by the induction hypothesis equals $N'$.

To show that $N$ equals $N^k$, consider the network $N'$ and apply the reverse of the operation that removed the
leaves $s_{5i-1},s_{5i-2},s_{5i},s_{5i-3}$ and $s_{5i-4}$ from $N$. This process is illustrated in
Fig.~\ref{fig:levelkladder}, and we will show that the such obtained network will equal $N^k$. Process the leaves in
reverse order, so add $s_{5i-4}$ to $N'$ first. Since $N'$ has $k-1$ reticulation leaves and $s_{5i}$ also has to
become a reticulation leaf, $s_{5i-4}$ must be a leaf below a split vertex. Hence $s_{5i-4}$ is added to the network
by hanging $s_{5i-4}$ on some arc of $N'$. The same holds for $s_{5i-3}$. Since $s_{5i}$ was a reticulation leaf in
$N$, it is added to the network choosing two arcs $(u_1,v_1),(u_2,v_2)$, subdividing them into $(u_1,w_1),(w_1,v_1)$
and $(u_2,w_2),(w_2,v_2)$, respectively, and adding a new reticulation vertex $x$ and arcs
$(w_1,x),(w_2,x),(x,s_{5i})$. Subsequently, $s_{5i-2}$ and $s_{5i-1}$ are added to the network by hanging them on arcs
to be specified. It remains to determine which arcs to subdivide, as to add the leaves
$s_{5i-1},s_{5i-2},s_{5i},s_{5i-3}$ and $s_{5i-4}$.

First consider the case $i > 1$. Because $s_{5k-4}s_{5i+1}|s_{5i-4}$ and $s_{5i-4}s_{5i+1}|s_{5i-7}$ are triplets in
$T^k$, it follows that $s_{5i-4}$ is added to $N'$ by hanging it on the arc entering the parent of $s_{5i+1}$.
Symmetrically, $s_{5i-3}$ is hung on the arc entering the parent of $s_{5i+2}$. This leads to network $N''$ in
Fig.~\ref{fig:levelkladder}. Next we discuss how to add $s_{5i}$ to network $N''$. Triplets $s_{5i}s_{5i+1}|s_{5i-4}$
and $s_{5k-4}s_{5i+1}|s_{5i}$ force a subdivision of the arc between the parents of $s_{5i-4}$ and $s_{5i+1}$. For
symmetric reasons, also the arc between the parents of $s_{5i+2}$ and $s_{5i-3}$ has to be subdivided. So subdivide
these arcs and make $s_{5i}$ a reticulation leaf below them (as described in detail in the previous paragraph). This
leads to the network $N'''$ in Fig.~\ref{fig:levelkladder}. Now $s_{5i-2}$ and $s_{5i-1}$ can only be added to the
network by hanging them on the arcs entering the parent of $s_{5i}$, since
$s_{5i+1}s_{5i-2}|s_{5i},s_{5i}s_{5i-2}|s_{5i+1}\in T^k$ and $s_{5i+2}s_{5i-1}|s_{5i},s_{5i}s_{5i-1}|s_{5i+2}\in T^k$.
This leads to the leftmost network in Fig.~\ref{fig:levelkladder}, which is the network $N^k$.

The case $i = 1$ is slightly different, since a leaf $s_{5i-7}$ does not exist. However, the triplets
$s_{5k-4}s_6|s_1$ and $s_6s_1|s_7$ enforce that $s_1 = s_{5i-4}$ is added to $N'$ by hanging it on the arc entering
the parent of $s_6$. Symmetrically, $s_2 = s_{5i-3}$ must be hung on the arc entering the parent of $s_7$. The same
arguments as in the case $i > 1$ show how to add the leaves $s_{5i},s_{5i-1},s_{5i-2}$. Also in this case we obtain
the network $N^k$.

It follows that $N$ equals $N^k$, completing the proof of Theorem~~\ref{thm:unique}.
\hfill$\Box$
\end{proof}

\begin{figure}
  \centering
  \begin{subfigure}[]
  {
    \centering
    \hspace{0.1cm}
    \includegraphics[scale=0.8]{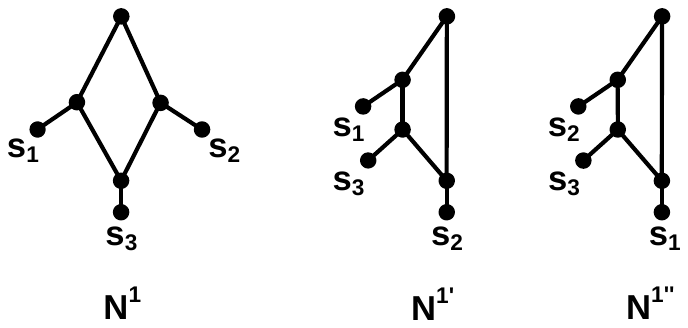}
    \hspace{0.1cm}
    \label{fig:uniquelevel1a}
  }
  \end{subfigure}
  \begin{subfigure}[]
  {
    \centering
    \hspace{0.9cm}
    \includegraphics[scale=0.8]{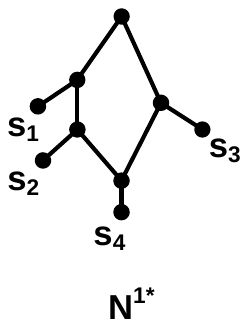}
    \hspace{0.9cm}
    \label{fig:uniquelevel1b}
  }
  \end{subfigure}
  \caption{(a) The three networks $N^1,{N^1}'$ and ${N^1}''$ that are consistent with $T^1 = \{s_1s_3|s_2,s_2s_3|s_1\}$ and (b) network ${N^1}^*$, which is the unique network that is consistent with the set of triplets consistent with ${N^1}^*$.}
  \label{fig:uniquelevel1}
\end{figure}

\smallskip
For level 1, the network $N^1$ is not the only network consistent with $T^1$.
Fig.~\ref{fig:uniquelevel1a} shows the three networks that are consistent with $T^1$.
However, there does exist a level-1 network that is unique in this sense. It is not too difficult to argue that the
network ${N^1}^*$ in Fig.~\ref{fig:uniquelevel1b} is the only level-1 network that is consistent with all triplets
that are consistent with ${N^1}^*$. For level 0, the only tree consistent with a single triplet is the triplet itself.

\section{From Uniqueness to Intractability of Constructing Level-$k$ Networks}\label{sec:nphard}
In this section we show how to use the unique networks from the previous section in the complexity analysis of network
reconstruction methods from triplets. We demonstrate this in two NP-hardness proofs. First, we show that it is
NP-hard, for each $k\geq 1$, to decide whether a given triplet set is consistent with some level-$k$ network.
Secondly, we show that the maximisation variant of this problem is NP-hard for each $k\geq 0$ even for dense triplet
sets.

We start with the proof that it is NP-hard to construct a level-$k$ network consistent with all input triplets.
Hardness was already known for $k = 1$ \cite{JanssonEtAl2006} and $k = 2$ \cite{VanIerselEtAl2007}. Note that the
uniqueness result from the previous section plays a crucial role in the subsequent NP-hardness proof for level $k$,
and that the NP-hardness is not a consequence of the hardness for levels 1 and 2.

In the proofs, we will often say we ``hang'' a leaf or
``caterpillar'' from a ``side'' of a simple level-$k$ generator. A
network is a \emph{caterpillar} if deleting all leaves gives a
directed path. In simple level-$k$ generators, a \emph{side} is
either an arc or a vertex with outdegree zero (cf.
\cite{VanIerselEtAl2007}). \emph{Hanging a caterpillar from arc}
$S_i$ means subdividing $S_i$ and connecting the new vertex to the
root of the caterpillar. Similarly defined is \emph{hanging a
caterpillar from a vertex with outdegree zero}, which gets connected
to the root of the caterpillar. Hanging a leaf from a side is
defined similarly. In addition, a leaf $x$ \emph{is on side} $S_i$
if there exists a cut-arc $(u,v)$ such that $u$ is on a subdivision
of $S_i$ (if $S_i$ is an arc) or $u$ is a reticulation vertex (if
$S_i$ is a reticulation vertex), and there is a directed path from
$v$ to $x$ (possibly $v = x$). A leaf $x$ is said to \emph{hang
between} vertices $w$ and $q$ if there is a cut-arc $(u,v)$ such
that $u$ is on a directed path from $w$ to $q$ and there is a
directed path from $v$ to $x$.

\begin{theorem}\label{thm:nphard}
For each $k\geq 2$, it is NP-hard to decide whether for a triplet set $T$ there exists some level-$k$ network $N$ consistent with $T$.
\end{theorem}
\begin{proof}
Reduce from the following NP-hard problem \cite{GareyJohnson1979}:\\
\\
\begin{tabular}{lp{0.85\textwidth}}
\multicolumn{2}{l}{\textsc{Set Splitting}}                                                                                                                                     \\
\textit{Instance:}  & A set $U = \{u_1,\hdots,u_n\}$ and a collection $\mathcal C = \{C_1,\hdots,C_m\}$ of size-3 subsets of $U$.                                              \\
\textit{Question:}  & Can $U$ be partitioned into sets $U_1$ and $U_2$ (a set splitting) such that $C_j\not\subseteq U_1$ and $C_j\not\subseteq U_2$, for all $1\leq j\leq m$? \\
                    & \\
\end{tabular}

From an instance $(U,\mathcal C)$ of \textsc{Set Splitting} construct a set $T$ of triplets as follows.
Start with triplet set $T^k$ (see previous Section), and for each set $C_j = \{u_a,u_b,u_c\}\in \mathcal C$ (with $1 \leq a < b < c \leq n$) add triplets $u_a^js_5|u_b^j$, $u_b^js_5|u_c^j$ and $u_c^js_5|u_a^j$.
In addition, for every $u_i\in U$ and $1\leq j\leq m$ add triplets $s_5u_i^j|s_1$, $s_5u_i^j|s_2$, $s_5s_6|u_i^j$, $s_5s_7|u_i^j$ and (if $j\neq m$) $u_i^ju_i^{j+1}|s_5$.
This completes the construction of $T$.
We will prove that $T$ is consistent with some level-$k$ network if and only if there exists a set splitting $\{U_1,U_2\}$ of $(U,\mathcal C)$.

First suppose that there exists a set splitting $\{U_1,U_2\}$ of $(U,\mathcal C)$.
Construct the network $N$ by starting with the network $N^k$, which is obtained from the simple level-$k$ generator $G^k$ in Fig.~\ref{fig:simplelevelk} by hanging a leaf $s_i$ on each side $S_i$.
\begin{figure}
  \centering
  \begin{subfigure}[The simple level-$k$ generator $G^k$.]
  {
    \centering
    \hspace{1.5cm}
    \includegraphics[scale=0.8]{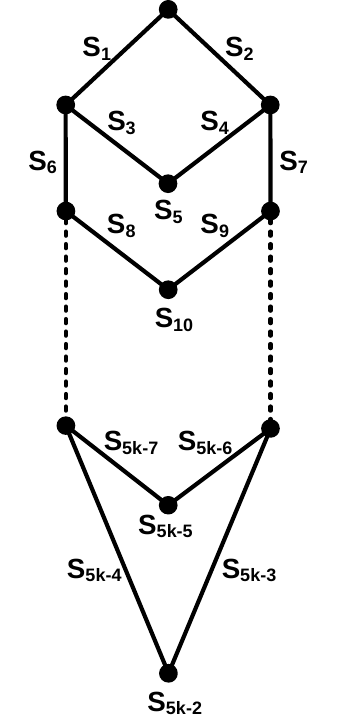}
    \hspace{1.5cm}
    \label{fig:simplelevelka}
  }
  \end{subfigure}
  \begin{subfigure}[The network $N$.]
  {
    \centering
    \includegraphics[scale=0.8]{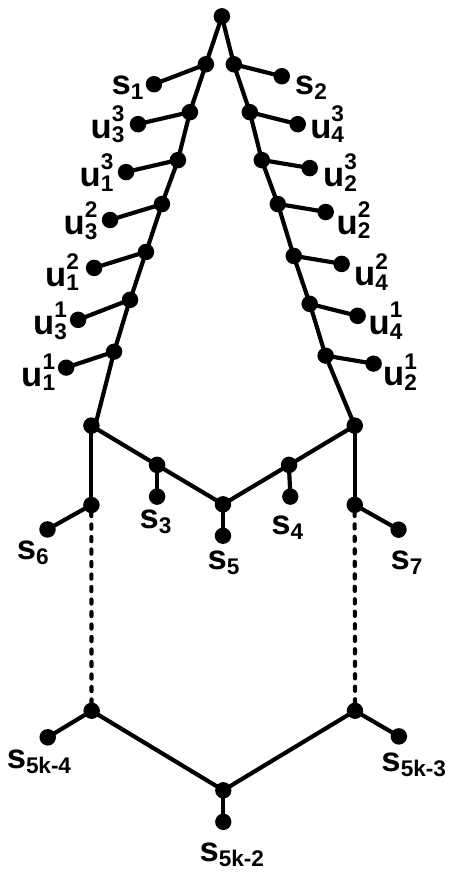}
  }
  \end{subfigure}
  \caption{Auxiliary networks in the proof of Theorem~\ref{thm:nphard}.}
  \label{fig:simplelevelk}
\end{figure}
For each element $u_i\in U_1$, hang all leaves $u_i^1,\hdots,u_i^m$ on side $S_1$ below the parent of $s_1$; for each element $u_i\in U_2$ hang all leaves $u_i^1,\hdots,u_i^m$ on side $S_2$ below the parent of $s_2$.
To determine the order in which to put these leaves consider a set $C_j = \{u_a,u_b,u_c\}\in \mathcal C$.
If $u_a$ and $u_b$ are in the same class of the partition, then put leaf $u_a^j$ below $u_b^j$; if $u_b$ and $u_c$ are in the same class of the partition put $u_b^j$ below $u_c^j$; and if $u_a$ and $u_c$ are in the same class put $u_c^j$ below $u_a^j$.
The rest of the ordering is arbitrary.
It is easy to check that $N$ is consistent with all triplets in $T$.
For an example of this construction see the network to the right in Fig.~\ref{fig:simplelevelk}.

Conversely, suppose that $T$ is consistent with some level-$k$ network $N$.
Since $T^k\subset T$, Theorem~\ref{thm:unique} says that $N$ must be equal to $N^k$ with the leaves not in $L(N^k)$ added.
Triplets $s_5u_i^j|s_1$ and $s_5u_i^j|s_2$ imply that none of the leaves $u_i^j$ can hang between the root and $s_1$, or between the root and $s_2$.
Further, triplets $s_5s_6|u_i^j$ and $s_5s_7|u_i^j$ imply that $u_i^j$ must be on either side $S_1$ or $S_2$.
Triplets $u_i^ju_i^{j+1}|s_5$ yield that for each $1\leq i\leq n$, all leaves $u_i^1,\hdots,u_i^m$ have to hang on the same side.
For $h\in\{1,2\}$, let $U_h$ be the set of elements $u_i\in U$ for which all leaves $u_i^1,\hdots,u_i^m$ hang on side $S_h$.
It remains to prove that $(U_1,U_2)$ is a set splitting of $(U,\mathcal C)$.
Consider a set $C_j = \{u_a,u_b,u_c\}$ and suppose for contradiction that $u_a,u_b,u_c\in U_h$ for some $h\in\{1,2\}$.
It follows that all leaves $u_a^j,u_b^j,u_c^j$ hang between $s_h$ and the root.
This is impossible, as $T$ contains triplets $u_a^js_5|u_b^j, u_b^js_5|u_c^j, u_c^js_5|u_a^j$.
\end{proof}

For dense triplet sets, it can be decided in polynomial time whether
there exists a level-1 \cite{JanssonEtAl2006} or level-2
\cite{VanIerselEtAl2008,VanIerselEtAl2007} network consistent with
all input triplets. Using the uniqueness result from the previous
section, we will prove that the maximization versions of these
problems are NP-hard, even for dense triplet sets and for all $k\geq
0$.

\noindent \begin{tabular}{lp{0.85\textwidth}}
\multicolumn{2}{l}{\textsc{MaxCL-$k$-Dense}}                                                                                           \\
\textit{Instance:}  & A dense triplet set $T$.                                                                                         \\
\textit{Output:}    & A level-$k$ network consistent with the maximum number of triplets in $T$ that any level-$k$ network is consistent with. \\
                    & \\
\end{tabular}

\begin{theorem}
The problem \textsc{MaxCL-$k$-Dense} is NP-hard, for all $k\geq 0$.
\label{thm:densehard}
\end{theorem}
\begin{proof}
Reduce from the following NP-hard problem \cite{Alon2006,CharbitEtAl2007}.\\
\\
\begin{tabular}{lp{0.85\textwidth}}
\multicolumn{2}{l}{\textsc{Feedback Arc Set in Tournaments (FAST)}}                                              \\
\textit{Instance:}  & A complete directed graph $G = (V,A)$ and an integer $q \in \mathbb{N}$.                   \\
\textit{Question:}  & Is there a set $A'\subseteq A$ of $q$ arcs such that $G' = (V, A\setminus A')$ is acyclic? \\
                    & \\
\end{tabular}

For $k = 0$, we imitate the reduction of the non-dense case \cite{Bryant1997,Wu2004}. The difference is that the
constructed instance of \textsc{MaxCL-0-Dense} contains more triplets, to become dense. Given an instance $G = (V,A)$
and $q\in \mathbb{N}$ of FAST, construct an instance $T$ of \textsc{MaxCL-0-Dense} as follows. Introduce a vertex $x
\not \in V$ and for each arc $(z,y)\in A$, add a triplet $xy|z$ to $T$. In addition, for each combination of three
leaves $v_1,v_2,v_3\in V$ (thus $v_1,v_2,v_3\neq x$), add all three triplets $v_1v_2|v_3$, $v_1v_3|v_2$ and
$v_2v_3|v_1$ to $T$. The differences with the reduction of Wu \cite{Wu2004} are that (1) we reduce from FAST instead
of FAS and that (2) we add all triplets containing three leaves from $V$. The combination of these two modifications
makes the instances dense. The extra triplets do not change the reduction since any level-0 network is consistent with
exactly one triplet for every combination of three leaves. The intuition of the reduction is as follows: the vertices
of an acyclic graph can be uniquely labeled such that arcs point only from vertices with higher label to vertices with
lower label. In a phylogenetic tree, this ordering of the vertices corresponds to an ordering of the leaves on the
unique path from the tree root to leaf $x$. Along the lines of the proof for the non-dense case
\cite{Bryant1997,Wu2004}, it can be argued that $G$ contains a feedback arc set of size $q$ if and only if there
exists a tree consistent with $|T| - q - 2\binom{|V|}{3}$ triplets from $T$. This completes the proof that
\textsc{MaxCL-0} is NP-hard for dense triplet sets.

For $k\geq 2$, use a similar reduction but start from the simple level-$k$ generator $G^k$ in
Fig.~\ref{fig:simplelevelka}. Use the following property of $G^k$, implied by Theorem~\ref{thm:unique}: Let $N^k$ be a
network obtained by hanging a leaf from each side of $G^k$. If $T^k$ is the triplet set consistent with $N^k$, then
$N^k$ is the unique level-$k$ network consistent with $T^k$.

Given a tournament $G = (V,A)$ and integer $q \in \mathbb{N}$, construct a corresponding instance $T$ of
\textsc{MaxCL-$k$-Dense} as follows. First construct a network $N'$ from $G^k$. From each side $S_i$ of $G^k$ hang a
caterpillar with leaves $S_i^1,\hdots,S_i^p$, with $p = 2(q + 2\binom{|V|}{3}) + 1$. The intuition being that $p$ is
``large'' to force a specific structure of the networks consistent with many triplets in $T$. For simplicity denote
$S_{5k-2}^1$ by $x$. Hang $|V|$ leaves on side $S_{5k-4}$, distinctly labeled by the vertices of $V$, between the root
of the caterpillar and the reticulation vertex on that side, in arbitrary order. This gives the network $N'$. For an
example, see the network on the right in Fig.~\ref{fig:densehard}. Let $T'$ be the set of triplets consistent with
$N'$, except for triplets $ab|c$ with $a,c\in V$ and $b\notin V$. For each arc $(z,y)\in A$, add a triplet $xy|z$ to
$T'$, informally encoding the arc $(z,y)$ as a constraint ``$z$ hangs between the root of the caterpillar and $y$''.
Finally, for each 3-set of vertices from $V$ add all three triplets over the three leaves labeled by the vertices,
that are not yet present. Denote the resulting (dense) triplet set by $T$, which forms an instance of
\textsc{MaxCL-$k$-Dense}. We will show that there exists a level-$k$ network $N$ consistent with $|T| - q -
2\binom{|V|}{3}$ triplets from $T$ if and only there exists a feedback arc set $A'$ of size $q$.

First suppose $G$ has a feedback arc set $A'$ of size $q$.
Thus the graph $G' = (V,A\setminus A')$ is acyclic, and each vertex $v\in V$ can receive a label $f(v)$ such that there are no arcs $(z,y)\in A\setminus A'$ with $f(y)\geq f(z)$.
Construct the network $N$ from $N'$ by rearranging the leaves from $V$ by sorting them with respect to their labels such that the highest leaf has the largest label.
For any arc $(z,y)\in A\setminus A'$ it holds that $f(y) < f(z)$ and hence the triplet $xy|z$ is consistent with $N$.
For every vertex pair $\{z,y\}$, the triplet $yz|x$ is consistent with $N$.
For each combination of three leaves from $V$ there is exactly one triplet over these leaves consistent with $N$.
It follows that the only triplets in $T$ that are not consistent with $N$ are (1) the triplets corresponding to the arcs in $A'$, and (2) exactly two-thirds of the triplets that have only leaves in $V$.
That means that in total $|T| - q - 2\binom{|V|}{3}$ triplets from $T$ are consistent with $N$.

For the converse, suppose there exists some level-$k$ network $N$ consistent with $|T| - q - 2\binom{|V|}{3}$ triplets from $T$.
For all $1\leq j\leq p$, there exists a unique network with leaf set $L_j = \{S_i^j~|~1\leq i\leq 5k-2\}$ that is consistent with all triplets from $T_j = T|_{L_j}$.
There are at most $q + 2\binom{|V|}{3}$ triplets not consistent with $N$, and the sets $T_j$ are pairwise disjoint, so at least one of the sets $L_j$ is placed on a simple level-$k$ network of type $G^k$.
Take any $i$ and observe that for each $j$ such that $S_i^j$ is not on side $S_i$ of $N$, there exists a triplet $t\in T$ that is not consistent with $N$ and
$L(t)=\{S_i^j,\ell_1,\ell_2\}$ for $\ell_1,\ell_2\not\in\{S_i^1,\hdots,S_i^p\}$.
If there would be more than $q + 2\binom{|V|}{3}$ such $j$ then there would be more than $q + 2\binom{|V|}{3}$ distinct triplets from $T$ not consistent with $N$.
Hence for each $i$ there are at least $p' = q + 2\binom{|V|}{3} + 1$ indices $j$ such that $S_i^j$ is on side $S_i$.
Let $L^*_1,\hdots,L^*_{p'}$ be pairwise disjoint sets each containing exactly one leaf $S_i^j$ that is on side $S_i$, for each $1\leq i \leq 5k-2$.

The next claim is that all leaves labeled by vertices from $V$ have to be on side $S_{5k-4}$, between the root of the caterpillar and the reticulation vertex on that side.
Suppose for contradiction this were not the case for some leaf labeled by $v\in V$.
Then for each of the leaf sets $L^*_1\cup\{v\},\hdots,L^*_{p'}\cup\{v\}$ there exists a triplet in $T$ not consistent with $N$.
Since the sets $L^*_1,\hdots,L^*_{p'}$ are pairwise disjoint and $p' > q + 2\binom{|V|}{3}$, we obtain a contradiction.

Since the leaves corresponding to vertices from $V$ all hang on the same side $S_{5k-4}$,
they can be uniquely labeled by their order on side $S_{5k-4}$, such that the highest leaf has the largest label.
If some leaves are below the same cut-arc, they receive the same label.
Let $A'$ be the set of arcs $(z,y)$ corresponding to the triplets $xy|z$ that are not consistent with $N$, and for every $v\in V$ let $f(v)$ be the label of the leaf corresponding to $v$.
Then the graph $G' = (V,A\setminus A')$ is acyclic, because all arcs $(z,y)\in A\setminus A'$ satisfy the relation $f(y) < f(z)$.

An example for $k = 2$ is displayed in Fig.~\ref{fig:densehard}.
\begin{figure}
  \centering
  \vspace{-.2cm}
  \includegraphics[scale=0.8]{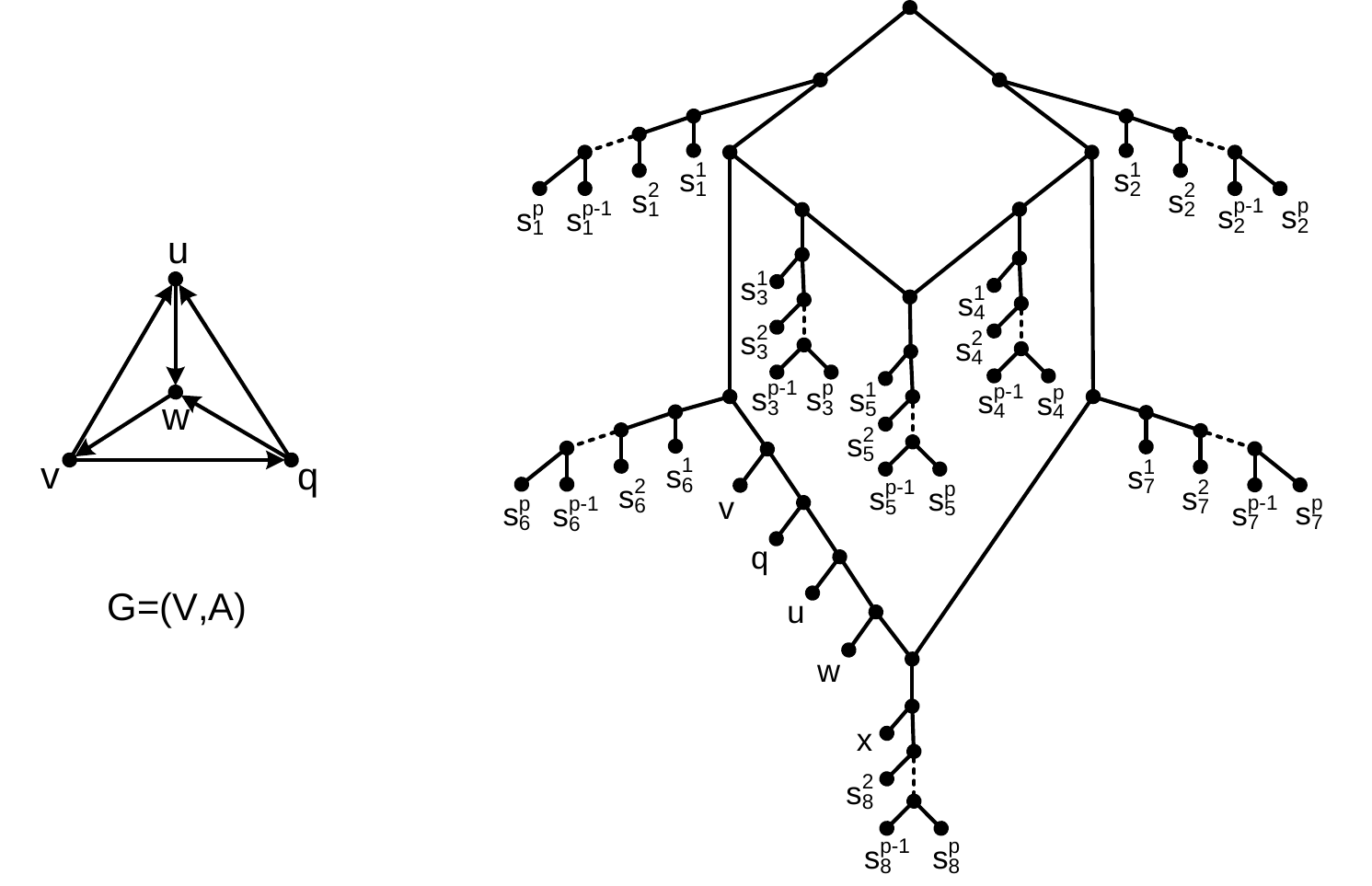}
  \caption{An example input $G = (V,A)$ of FAST on the left and the network $N$ constructed in the proof of Theorem~\ref{thm:densehard}, for $k=2$, to the right.}
  \vspace{-0.5cm}
  \label{fig:densehard}
\end{figure}
The graph on the left is an example instance $G = (V,A)$ of FAST.
The arcs of $G$ are encoded as triplets $xw|u$, $xw|q$, $xu|v$, $xv|w$, $xu|q$ and $xq|v$.
The network $N$ to the right is consistent with all these triplets except $xv|w$.
The arc $(w,v)$ is indeed a feedback arc set of the graph $G$.
Other triplets in $T$ enforce this specific level-2 network $N$ and make $T$ dense.\\
\\
For $k = 1$ the same reduction as for $k\geq 2$ works, when hanging two caterpillars from side $S_1$.
\end{proof}

\section{An Exact Algorithm for Constructing Level-1 Networks}
\label{sec:exact} Given the intractability results from the previous section for constructing networks consistent with
a maximum number of input triplets, there is no hope (unless $\textnormal{P} = \textnormal{NP}$) for algorithms
solving these problems exactly and in polynomial time. Still, these problems need to be solved in practice, so
algorithms for \textsc{MaxCL-$k$-Dense} and its relaxation \textsc{MaxCL-$k$} to general triplet sets are either not
guaranteed to give an optimal solution, or require superpolynomial time. In this section we consider the latter
approach.

Wu described an exact algorithm \cite{Wu2004} that finds a tree consistent with a maximum number of input triplets in
$O(3^n(n^2+m))$ time, with $m$ the number of triplets and $n$ the number of leaves. We extend this approach for
reconstructing evolutions that are not tree-like, but where reticulation cycles are disjoint. We do this by describing
an exact algorithm that runs in $O(m4^n)$ time and solves the \textsc{MaxCL-1} problem, which is NP-hard by
Theorem~\ref{thm:densehard}.

Note that the problem \textsc{MaxCL-1} does not only ask if there exists a level-1 network consistent with \emph{all}
input triplets; it asks us to find a level-1 network that is consistent with a maximum number of them. Hence an
algorithm for \textsc{MaxCL-1} always outputs a solution, no matter how bad the data is the algorithm is confronted
with. This contrasts with existing algorithms \cite{AhoEtAl1981,VanIerselEtAl2008,JanssonSung2006} that only find a
solution if a network exists that is consistent with all triplets of a (dense) input. The algorithm described in this
section is also more powerful in that it also works for non-dense triplet sets. It can thus be used even if for some
combinations of three taxa it is difficult to find the right triplet, which is very likely to be the case in practice.
The very same algorithm works for the weighted version of the problem. In addition, it can also be used to choose,
among all level-1 networks consistent with a maximum number of input triplets, a network with a minimum number of
reticulation vertices. However, its exponential running time means that it can only be used for a relatively small
number of leaves at a time.

The intuition behind our algorithm is the following. There are three different shapes possible for the optimal
network. Either the arcs leaving the root are cut-arcs, like in Fig.~\ref{fig:exact1b}, or the root is part of a
cycle, which can be ``skew'' like the cycle in Fig.~\ref{fig:exact2a} or ``non-skew'' like in Fig.~\ref{fig:exact2b}.
We can try to construct a network of each type separately. Given the tripartition $(X',Y',Z')$ or bipartition
$(X',Y')$ of the leaves indicated in the figures, it turns out to be possible to reconstruct the optimal network by
combining optimal smaller networks for $X'$, $Y'$, $X'\cup Z'$ and $Y'\cup Z'$. Critical is that these smaller
networks for $X'\cup Z'$ and $Y'\cup Z'$ must be such that combining the different networks does not create
biconnected components with more than one reticulation vertex.

To achieve this, we introduce the notion of
``non-cycle-reachable''-arc, or n.c.r.-arc for short. An arc $a =
(u,v)$ is an \emph{n.c.r.-arc} if there is no directed path of
length at least one from any vertex $w$ in an (undirected) cycle to
$u$. Also, for some arc $a = (u,v)$ write $R[a]$ to denote the set
of leaves below $v$. Use $f_T(N)$ to denote the number of triplets
in $T$ consistent with $N$; and $g_T(N,Z)$ to denote the number of
triplets in $T$ consistent with $N$ and that are not of the form
$xy|z$ with $z\in Z$ and $x,y\notin Z$. Write $f(N)$ as short for
$f_T(N)$ and $g(N,Z)$ for $g_T(N,Z)$. It will become clear later
that the definition of $g$ ensures that combining networks that are
optimal w.r.t. $g$ leads to networks optimal w.r.t. $f$.

The algorithm works as follows.
Loop through all subsets $L'\subseteq L$ in increasing cardinality and consider each tripartition $\pi(L') = (X,Y,Z)$ with $X,Y\neq\emptyset$.
While the r\^{o}les of $X$ and $Y$ are symmetric, this is not the case for $X$ and $Z$, and $Y$ and $Z$.
The following networks have been computed in previous iterations of the algorithm:
\begin{itemize}
  \item A network $N^{X}$ maximizing $f(N)$ over all level-1 networks $N$ with $L(N) = X$;
  \item A network $N^{Y}$ maximizing $f(N)$ over all level-1 networks $N$ with $L(N) = Y$;
  \item A network $N^{XZ}$ maximizing $g(N,Z)$ over all level-1 networks $N$ with $L(N) = X\cup Z$ that contain an n.c.r.-arc $a$ with $Z=R[a]$;
  \item a network $N^{YZ}$ maximizing $g(N,Z)$ over all level-1 networks $N$ with $L(N) = Y\cup Z$ that contain an n.c.r.-arc $a$ with $Z=R[a]$.
\end{itemize}
If $Z=\emptyset$, combine $N^{X}$ and $N^{Y}$ into a new network $N^2_\pi$ by adding a new root and connecting it to
the roots of $N^{X}$ and $N^{Y}$. If $Z\neq\emptyset$, proceed as follows:
\begin{enumerate}
\item Combine $N^{XZ}$ and $N^{YZ}$ into a new network $N^1_\pi$ by creating a ``non-skew'' cycle as follows.
      Add a new root and connect it to the roots of $N^{XZ}$ and $N^{YZ}$.
      Let $a = (u,v)$ and $a' = (u',v')$ be the (unique) n.c.r.-arcs such that $Z = R[a]$ in $N^{XZ}$ and $Z = R[a']$ in $N^{YZ}$.
      Subdivide $a$ into $(u,w)$ and $(w,v)$, delete $v'$ and all arcs and vertices reachable from $v'$, and add an arc $(u',w)$ (see Fig.~\ref{fig:algorithm1});
      \begin{figure}
      \centering
      \includegraphics[scale=0.8]{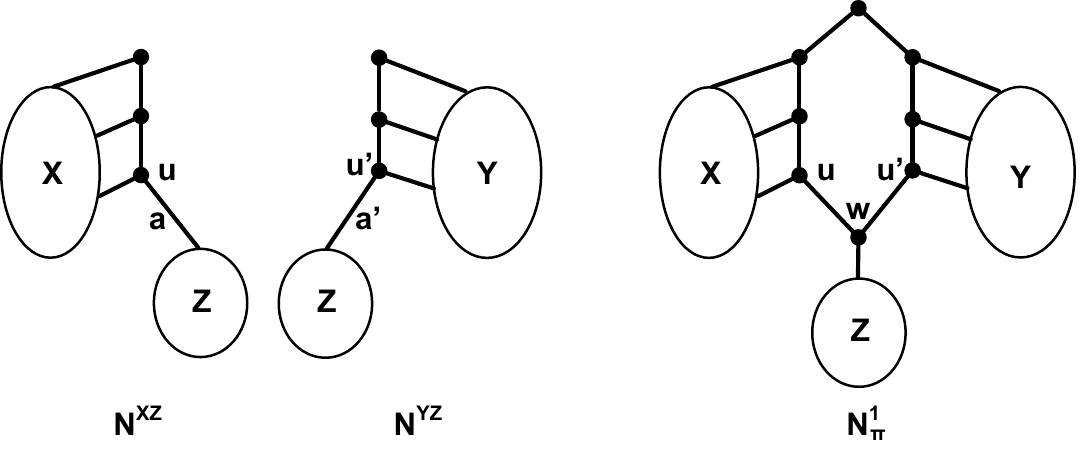}
      \caption{Construction of $N_\pi^1$ from $N^{XZ}$ and $N^{YZ}$.}
      \label{fig:algorithm1}
      \end{figure}
\item Combine $N^{X}$ and $N^{YZ}$ into a new network $N^2_\pi$ by adding a new root and connecting it to the roots of $N^{X}$ and $N^{YZ}$;
\item Create $N^3_\pi$ from $N^2_\pi$ by creating a ``skew'' cycle as follows: let $a = (u,v)$ be the (unique) n.c.r.-arc with $Z = R[a]$.
      Subdivide $a$ into $(u,w)$ and $(w,v)$, add a new root and connect it to the old root and to $w$ (see Fig.~\ref{fig:algorithm2}).
      \begin{figure}
      \centering
      \includegraphics[scale=.8]{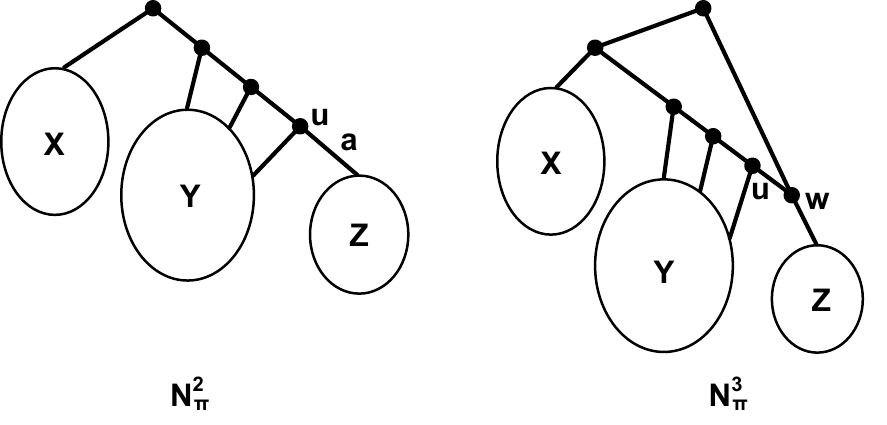}
      \caption{Networks $N_{\pi}^2$ and $N_{\pi}^3$ if $Z\neq\emptyset$.}
      \label{fig:algorithm2}
     \end{figure}
\end{enumerate}

Let $N(L')$ be a network that maximizes $f(N)$ over the networks
$N^1_\pi,N^2_\pi$ and $N^3_\pi$ over all tripartitions $\pi(L')$. In
addition, for each $\bar{Z}\subset L'$, let $N^2(L',\bar{Z})$ be a
network that maximizes $g(N^2_\pi,\bar{Z})$ over the networks
$N^2_\pi$ over all tripartitions $\pi(L') = (X,Y,Z)$ with
$Z=\bar{Z}$. This concludes the description of the algorithm.

Because the arcs $a=(u,v)$ and $a'=(u',v')$ in steps (1) and (3) are
n.c.r.-arcs, we know that (in $N^{XZ}$ and $N^{YZ}$) neither $u$,
nor $u'$, nor one of their ancestors is contained in a cycle. It
follows that the newly created cycles do not overlap with any of the
original cycles and hence that the constructed networks are indeed
level-1 networks. It now also becomes clear why networks $N^{XZ}$
and $N^{YZ}$ are used that are optimal w.r.t. $g$ (rather than $f$).
The creation of a new cycle, as in Fig.~\ref{fig:algorithm1} and
Fig.~\ref{fig:algorithm2}, causes all triplets of the form $xy|z$
with $z\in Z$ and $x,y\notin Z$ to become consistent with the
network.

We claim that $N(L')$ maximizes $f(N)$ over all level-1 networks $N$
with $L(N)=L'$. This implies that, in each iteration of the
algorithm, the networks $N^X$ and $N^Y$ have indeed been computed in
a previous iteration. This claim also implies that the algorithm
finds an optimal solution.

In addition, we claim that $N^2(L',\bar{Z})$ maximizes
$g(N,\bar{Z})$ over all level-1 networks $N$ with $L(N)=L'$ that
contain an n.c.r.-arc $a$ with $\bar{Z}=R[a]$. This implies that, in
each iteration, the networks $N^{XZ}$ and $N^{YZ}$ have indeed been
computed in a previous iteration of the algorithm.

The above claims are proved by induction on the size of $L'$. They
do hold for sets $L'$ with $|L'|\leq 3$; so given some leaf set $L'$
with $|L'| > 3$ assume that the above statements hold for all leaf
sets of smaller size. We will show that the statements are then also
true for $L'$. Observe that from the induction hypothesis follows
that we may take $N(X)$ to be $N^X$, which has hence indeed be
computed in a previous iteration of the algorithm. Similarly, we may
take $N^2(X\cup Z,Z)$ and $N^2(Y\cup Z,Z)$ to be $N^{XZ}$ and
$N^{YZ}$, respectively, which have also indeed been computed in a
previous iteration of the algorithm. The induction step then follows
from the following two lemmas.

\begin{lemma}
For every $\bar{Z}\neq \emptyset$, the network $N^2(L',\bar{Z})$ maximizes $g(N,\bar{Z})$ over all level-1 networks
$N$ with $L(N) = L'$ that contain an n.c.r.-arc $a$ with $\bar{Z}=R[a]$.\label{lem:exact1}
\end{lemma}
\begin{proof}
Let $N'$ be a network with $L(N) = L'$ and some n.c.r.-arc $a$ such that $\bar{Z}=R[a]$.
We show that $g(N',\bar{Z}) \leq g(N^2(L',\bar{Z}),\bar{Z})$.
Because $N'$ contains the n.c.r.-arc $a$, the root of $N'$ is not in a cycle.
Let $a_1$ and $a_2$ be the two cut-arcs leaving the root such that the leaves in $\bar{Z}$ are reachable from $a_2$.
Let $X'=R[a_1]$ and $Y'=R[a_2]\setminus \bar{Z}$, see Fig.~\ref{fig:exact1a}.
Because $N^2(L',\bar{Z})$ maximizes $g(N_\pi^2,\bar{Z})$ over all tripartitions $(X,Y,Z)$ of $L'$ with $Z = \bar{Z}$, it is certainly at least as good as $N_{(X',Y',\bar{Z})}^2$.
That is, $g(N^2(L',\bar{Z}),\bar{Z}) \geq g(N_{(X',Y',\bar{Z})}^2,\bar{Z})$.
Write $N^2{}'$ as short for $N_{(X',Y',\bar{Z})}^2$.
Compare triplets consistent with $N'$, with those consistent with $N^2{}'$.
\begin{itemize}
\item There are at least as many triplets in $T|_{X'}$ consistent with $N^2{}'$ as with $N'$, because $N^{X'}$ is a subgraph of $N^2{}'$ and $N^{X'}$ maximizes $f(N)$ over all networks $N$ with $L(N)=X'$.
\item There are at least as many triplets in $T|_{(Y'\cup Z)}$ that are not of the form $y_1y_2|z$ for $y_1,y_2\in Y'$ and $z\in \bar{Z}$ that are consistent with $N^2{}'$ as with $N'$,
      because $N^{Y'\bar{Z}}$ is a subgraph of $N^2{}'$ and $N^{Y'\bar{Z}}$ maximizes $g(N,\bar{Z})$ over all networks $N$ with $L(N) = Y'\cup \bar{Z}$ that contain an n.c.r.-arc $a$ with $\bar{Z} = R[a]$.
\item All triplets of the form $ab|c$ with $a,b\in X'$, $c\in Y'\cup \bar{Z}$ or $a,b\in Y'\cup\bar{Z}$, $c\in X'$ are consistent with both $N^2{}'$ and $N'$.
\item All triplets of the form $ab|c$ with $a,c\in X'$, $b\in Y'\cup \bar{Z}$ or $a,c\in Y'\cup\bar{Z}$, $b\in X'$ are consistent with neither $N^2{}'$ nor $N'$.
\end{itemize}
Thus $g(N',\bar{Z}) \leq g(N^2{}',\bar{Z}) = g(N^2(L',\bar{Z}),\bar{Z})$.
\end{proof}

\begin{lemma}
The network $N(L')$ maximizes $f(N)$ over all level-1 networks $N$ with $L(N) = L'$.
\label{lem:maxconsensus}
\end{lemma}
\begin{proof}
For contradiction, suppose that some network $N' \not= N(L')$ with $L(N') = L'$ is consistent with more triplets in
$T$ than $N(L')$. Distinguish three cases, depending on the shape of $N'$.

The first case is that the two arcs leaving the root of $N'$ are cut-arcs $a_1$ and $a_2$.
Let $X'=R[a_1]$, $Y'=R[a_2]$ and $Z' = \emptyset$, see Fig.~\ref{fig:exact1b}, and compare $N'$ to $N^2_{(X',Y',Z')}$.

\begin{figure}
  \centering
  \begin{subfigure}[In the proof of Lemma~\ref{lem:exact1}.]
  {
    \centering
    \includegraphics[scale=0.9]{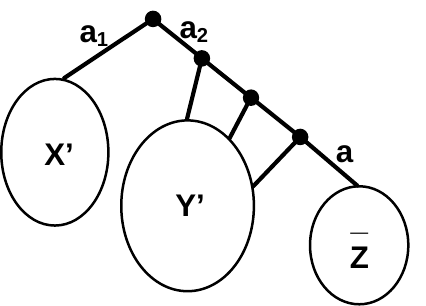}
    \label{fig:exact1a}
  }
  \end{subfigure}
  \begin{subfigure}[The first case in the proof of Lemma~\ref{lem:maxconsensus}.]
  {
    \centering
    \hspace{2cm}
    \includegraphics[scale=0.9]{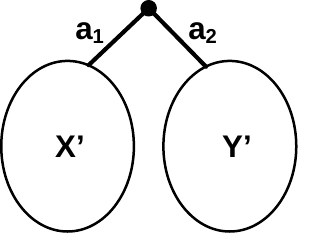}
    \hspace{2cm}
    \label{fig:exact1b}
  }
  \end{subfigure}
  \caption{Network $N'$ in the proofs of Lemmas~\ref{lem:exact1} and \ref{lem:maxconsensus}.}
\end{figure}

The latter network is consistent with at least as many triplets from $T|_{X'}$ because it contains $N^{X'}$ as a
subnetwork, and $N^{X'}$ maximizes $f(N)$ over all networks $N$ with $L(N) = X'$. Similarly, the network
$N^2_{(X',Y',Z')}$ is consistent with as least as many triplets from $T|_{Y'}$ as $N'$. All other triplets are either
consistent with both or with none of these networks. Hence $N^2_{(X',Y',Z')}$ is consistent with at least as many
triplets as $N'$. Because $N(L')$ is consistent with at least as many triplets as $N^2_{(X',Y',Z')}$, it follows that
$N(L')$ is also consistent with at least as many triplets as $N'$; a contradiction.

The second case is that one child of the root of $N'$ is a reticulation vertex. Let $a_1=(r,v_1)$ and $a_2=(r,v_2)$ be
the two arcs leaving the root of $N'$ and suppose that $v_2$ is a reticulation vertex. Let $a_3$ and $a_4$ be the two
arcs leaving $v_1$. Because $N'$ is a level-1 network, one of $a_3,a_4$ is a cut-arc, say $a_3$. Let $X' = R[a_3]$,
$Y' = R[a_4]\setminus R[a_2]$ and $Z' = R[a_2]$, see Fig.~\ref{fig:exact2a}. Compare the networks $N'$ and
$N^3_{(X',Y',Z')}$ with respect to the number of triplets in $T$ these networks are consistent with. First, consider
triplets in $T|_{X'}$: Network $N^3_{(X',Y',Z')}$ is consistent with at least as many of these as $N'$, because it
contains $N^{X'}$ as a subgraph. Second, consider triplets in $T|_{(Y'\cup Z')}$ that are not of the form $y_1y_2|z$
for $y_1,y_2\in Y$ and $z\in Z$. We will show that $N^3_{(X',Y',Z')}$ is consistent with at least as many of these
triplets as $N'$. First recall that $N^3_{(X',Y',Z')}$ contains a subdivision of $N^{Y'Z'}$, which maximizes $g(N,Z')$
over all networks with $L(N) = Y'\cup Z'$ containing an n.c.r.-arc $a$ with $Z' = R[a]$. The network $N'$ does not
contain such an n.c.r.-arc, but we will modify it to a network that does contain such an n.c.r.-arc and is consistent
with the same number of the considered triplets. Let $N''$ be the network $N'$ with the arc $a_2$ removed. Observe
that $g(N'',Z') = g(N',Z')$, and so it follows that $N^3_{(X',Y',Z')}$ is consistent with at least as many of the
considered triplets as $N'$. All other triplets are either consistent with both $N^3_{(X',Y',Z')},N'$ or with none,
since both networks have the structure from Fig.~\ref{fig:exact2a}: only the internal structure inside $X'$, $Y'$ and
$Z'$ might be different in the two networks. Hence $N^3_{(X',Y',Z')}$ is consistent with at least as many triplets as
$N'$. Because $N(L')$ is consistent with at least as many triplets as $N^3_{(X',Y',Z')}$ it follows that $N(L')$ is
also consistent with at least as many triplets as $N'$; a contradiction.

\begin{figure}
  \centering
  \begin{subfigure}[Skew cycle; the second case in Lemma~\ref{lem:maxconsensus}.]
  {
    \centering
    \includegraphics[scale=0.9]{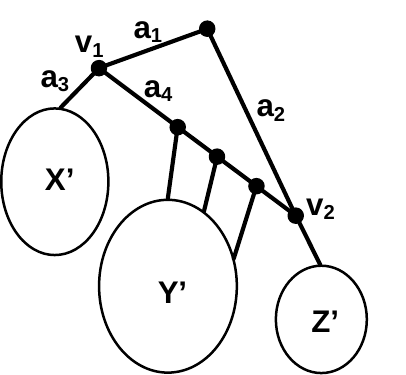}
    \label{fig:exact2a}
  }
  \end{subfigure}
  \begin{subfigure}[Non-skew cycle; the third case in Lemma~\ref{lem:maxconsensus}.]
  {
    \centering
    \includegraphics[scale=0.9]{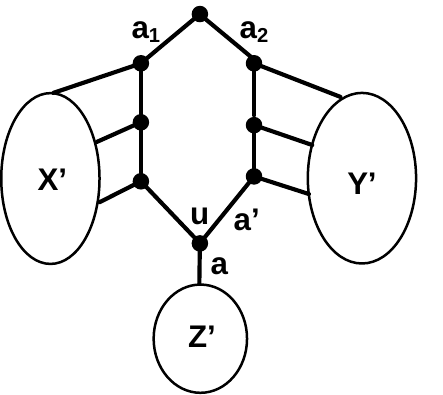}
    \label{fig:exact2b}
  }
  \end{subfigure}
  \caption{Shapes of networks, referred to as network $N'$ in the proof of Lemma~\ref{lem:maxconsensus}.}
\end{figure}

The last case is that the two arcs $a_1$ and $a_2$ leaving the root of $N'$ are not cut-arcs and are also not leading
to reticulation vertices. Let $X' = R[a_1]\setminus R[a_2]$, $Y' = R[a_2]\setminus R[a_1]$ and $Z' = R[a_1]\cap
R[a_2]$, see Fig.~\ref{fig:exact2b}. Compare the networks $N'$ and $N^1_{(X',Y',Z')}$, with respect to the number of
triplets in $T$ these networks are consistent with. First, consider triplets in $T|_{(X'\cup Z')}$ that are not of the
form $x_1x_2|z$ for $x_1,x_2\in X'$ and $z\in Z'$. We will show that $N^1_{(X',Y',Z')}$ is consistent with at least as
many of these triplets as $N'$. Recall that $N^1_{(X',Y',Z')}$ contains a subdivision of $N^{X'Z'}$, which maximizes
$g(N,Z')$ over all networks with $L(N) = X'\cup Z'$ containing an n.c.r.-arc $a$ with $Z' = R[a]$. The network $N'$
does not contain such an n.c.r.-arc, but we will modify it to a network that does contain such an n.c.r.-arc and is
consistent with the same number of the considered triplets. Let $a = (u,v)$ be the cut-arc in $N'$ with $Z' = R[a]$,
and let $a'$ be the arc that leads to $u$ and is reachable from $a_2$. Let $N''$ be the network $N'$ with the arc $a'$
removed. Now $N''$ is consistent with the same number of the considered triplets as $N'$, and so it follows that
$N^1_{(X',Y',Z')}$ is consistent with at least as many of the considered triplets as $N'$. In a similar way it follows
that $N^1_{(X',Y',Z')}$ is consistent with as least as many triplets in $T|_{(Y'\cup Z')}$ that are not of the form
$y_1y_2|z$ for $y_1,y_2\in Y'$ and $z\in Z'$). All other triplets are either consistent with both networks or with
none. Hence $N^1_{(X',Y',Z')}$ is consistent with at least as many triplets as $N'$. Because $N(L')$ is consistent
with at least as many triplets as $N^1_{(X',Y',Z')}$ it follows that $N(L')$ is also consistent with at least as many
triplets as $N'$; a contradiction.
\end{proof}

\begin{theorem}
Given a set $T$ of $m$ triplets over $n$ leaves, a level-1 network consistent with a maximum number of triplets in $T$ can be constructed in $O(m4^n)$ time and $O(n3^n)$ space.
\end{theorem}
\begin{proof}
To achieve a small polynomial factor in the complexity, we use dynamic programming to compute the optimal value of the
solution as well as the partitions we have to choose in each step. Then a traceback algorithm constructs a network
consistent with the maximum number of triplets. To be precise, the dynamic programming algorithm finds, for all
$L'\subseteq L$, the maximum number $\hat{f}(L')$ of triplets in $T$ consistent with a level-1 network with leaves
$L'\subseteq L$. It also computes, for all $\bar{Z}\subset L'$, the maximum value $\hat{g}(L',\bar{Z})$ of
$g(N,\bar{Z})$ over all level-1 networks $N$ with leaves $L'$ that contain an n.c.r.-arc $a$ with $\bar{Z}=R[a]$. The
algorithm loops through all the subsets $L'\subseteq L$ from small to large and considers all tripartitions $\pi(L') =
(X,Y,Z)$. For each such partition, the values $\hat{f}(X)$, $\hat{f}(Y)$, $\hat{f}(Z)$, $\hat{g}(X\cup Z,Z)$ and
$\hat{g}(Y \cup Z,Z)$ are readily available from previous iterations. To compute the values $\hat{f}(L')$ and
$\hat{g}(L',Z)$ it only remains to count certain triplets in $T$, whose consistency with a network only depends on the
tripartition $(X,Y,Z)$ and the network type ($N_\pi^1$, $N_\pi^2$ or $N_\pi^3$). This can be done by first checking
membership of $X$, $Y$ and $Z$ for each leaf in $L'$ (in $O(n)$ time) and then looping through all triplets only once.
Hence this counting can be done in $O(n+m) = O(m)$ time. The algorithm's overall running time is thus bounded by
$O(m)\sum_{\ell=1}^n\binom{n}{\ell}O\left(3^\ell\right) = O(m4^n)$.

For each leaf set $L'\subseteq L$, store the optimal tripartition and the optimal type of network ($N_\pi^1$, $N_\pi^2$ or $N_\pi^3$).
In addition, store an optimal bipartition for all $L'\subseteq L$ and $\bar{Z}\subset L'$.
This yields a total space complexity of $O(n3^n)$.

Once the values $\hat{f}(L')$ and $\hat{g}(L',\bar{Z})$ have been computed and all optimal tripartitions and
bipartitions have been stored, a level-1 network $N$ consistent with $\hat{f}(L)$ many triplets can be constructed by
traceback, in polynomial time. Optimality of the algorithm follows from
Lemmas~\ref{lem:exact1}~and~\ref{lem:maxconsensus}.
\end{proof}

\section{Open Problems}
The obvious question to ask is whether the $O(m4^n)$ running time of our exact algorithm in Sect.~\ref{sec:exact} can be improved.
The same question can be asked about the $O(3^n(n^2+m))$ algorithm \cite{Wu2004} for maximum consistent trees.
It would also be interesting to extend the exact approach to the construction of level-2 networks, provided that reasonable running times can be achieved.

Positive results for level-3 and higher networks have so far remained out of reach.
In light of 65 simple level-3 generators \cite{Kelk2008}, we fear that algorithms for constructing level-3 networks (and higher) will almost certainly not be possible by using approaches similar to the ones in Sect.~\ref{sec:exact}.
A similar statement holds for the dense level-2 case \cite{VanIerselEtAl2008}, since the devised algorithms explicitly distinguish between the structures of different level-$k$ generators.
Tantalisingly, however, it remains a possibility that for each $k\geq 0$ it is polynomial-time solvable to determine whether there is a level-$k$ network consistent with a dense set of input triplets.

Approximability of the \textsc{MaxCL-$k$} problem needs to be further explored. APX-completeness of \textsc{MaxCL-0}
is known \cite{ByrkaEtAl2008}, hence no \emph{Polynomial Time Approximation Scheme} for \textsc{MaxCL-0} is possible
unless $\textnormal{P} = \textnormal{NP}$. It would be interesting to extend this result to $k > 0$. On the other
hand, the best known approximation ratios are $\frac{1}{3}$ for \textsc{MaxCL-0} \cite{Gasieniec1999} and $0.48$ for
\textsc{MaxCL-1} \cite{ByrkaEtAl2008}, leaving (potentially) much room for improvement.

From a more practical point of view, it is worthwhile to study the actual level of real evolutionary histories.
This will tell for which values of $k$ it remains important to design algorithms that construct level-$k$ networks.

\bibliographystyle{abbrv}
\bibliography{reflections}

\end{document}